\documentclass[prd,showpacs,showkeys,nofootinbib,12pt,amsfonts]{revtex4}

\def\a{\alpha}
\def\b{\beta}
\def\g{\gamma}
\def\vta{\vartheta}
\def\d{\delta}

\begin{document}

\title{A remark on an ansatz by M.W.\ Evans and the so-called
  Einstein-Cartan-Evans unified field theory}

\author{Friedrich W.\
  Hehl\email{hehl@thp.uni-koeln.de}}

\affiliation{Institute for Theoretical Physics, University at Cologne,
  50923 K\"oln, Germany \\ and\\Department of Physics and Astronomy,
  University of Missouri-Columbia, Columbia, MO 65211, USA}

\date{2006-12-01, {\em file ECE1.tex}}


\begin{abstract}
  M.W.\ Evans tried to relate the electromagnetic field strength to
  the torsion of a Riemann-Cartan spacetime. We show that this ansatz
  is untenable for at least two reasons: (i) Geometry: Torsion is
  related to the (external) translation group and cannot be linked to
  an internal group, like the $U(1)$ group of electrodynamics. (ii)
  Electrodynamics: The electromagnetic field strength as a 2-form
  carries 6 independent components, whereas Evans' electromagnetic
  construct $F^\a$ is a vector-valued 2-form with 24 independent
  components. This doesn't match. One of these reasons is already
  enough to disprove the ansatz of Evans.
\end{abstract}

\pacs{03.50.Kk; 04.20.Jb; 04.50.+h}

\keywords{}

\maketitle


\section{Introduction}

In 2005, Evans \cite{Evans} related electromagnetism to the torsion of
spacetime. We came across this paper in the context of a refereeing
process. We immediately recognized that the ansatz of Evans is shaky;
in fact, it will turn out to be incorrect.

As a convenient starting point for our discussion, we can take the
viable Einstein-Cartan theory of gravity. There the torsion becomes
very transparent from a geometrical as well as from a physical point
of view (Sec.II). Then, we come to the Evans ansatz (Sec.III) and
show that it represents an overkill for the only 6 components of the
electromagnetic field strength. 

As a historical note we add that attempts in the direction of the
Evans ansatz started in 1925 by Eyraud \cite{Eyraud} and in 1926 by
Infeld \cite{Infeld} and were shown to lead to nowhere, see Goenner
\cite{Goenner} and Tonnelat \cite{Tonnelat}.

\section{Einstein-Cartan theory of gravity (ECT)}

It is known from the literature that the Einstein-Cartan theory 
of gravity, see, for instance, \cite{RMP,Trautman}, is a {\it viable}
theory of gravitation. Torsion, which geometrically is related to {\it
  translations} (it describes closure failures of infinitesimal
parallelograms), is, according to the second field equation of the
ECT, proportional to the spin angular momentum density of matter. This
is not an ansatz, but the result of a variational principle of the
Hilbert-Einstein type and of the Noether procedure identifying the
right-hand-sides of the field equations as energy-momentum and spin
angular momentum, respectively.

Let us shortly sketch the formalism, for details see \cite{Erice}. We
start form the coframe $\vta^\a$, the metric
$g=g_{\a\b}\,\vta^\a\!\otimes\vta^\b$, and the connection
$\Gamma_\a{}^\b$. Here $\a,\b,...=0,1,2,3$ are anholonomic or frame
indices. With the connection, we can define an exterior covariant
derivative $D$. In the ECT, one assumes vanishing nonmetricity
\begin{equation}\label{nonmetricity}
  Q_{\a\b}:=-Dg_{\a\b}=0\,.
\end{equation}
The torsion $T^\a$ and the curvature $R_\a{}^\b$ are defined by
\begin{eqnarray}\label{torsion}
  T^\a&:=&D\vta^\a =d\vta^\a+\Gamma_\b{}^\a\wedge \vta^\b\,,\\
  R_\a{}^\b&:=& d\Gamma_a{}^\b- \Gamma_\a{}^\g\wedge\Gamma_\g{}^\b\,.
\label{curvature}
\end{eqnarray}
These two quantities fulfill the first and the second Bianchi
identities:
\begin{eqnarray}\label{Bianchi1}
DT^\a&\equiv& R_\b{}^\a\wedge \vta^\b\,,\\
DR_a{}^\b&\equiv&0\,.\label{Bianchi2}
\end{eqnarray}

Both can be decomposed irreducibly under the local Lorentz group
\cite{HMcCrea}. We define the 1-form $\eta_{\a\b\g}:=\,^\star\!\left(
  \vta_\a\wedge\vta_\b\wedge\vta_\g\right)$, where the star denotes
the Hodge operator. From (\ref{Bianchi1}) we pick the irreducible
piece with 6 independent components,
\begin{equation}\label{Bianchi1irr}
  DT^\g\wedge\eta_{\g\a\b}\equiv
R_\d{}^\g\wedge\vta^\d\wedge \eta_{\g\a\b}\,,
\end{equation}
and from (\ref{Bianchi2}) one with 4 independent components,
\begin{equation}\label{Bianchi2irr}
DR^{\b\g}\wedge\eta_{\b\g\a}\equiv 0\,.
\end{equation}
We will come back to these equations below. The equations
(\ref{nonmetricity}) to (\ref{Bianchi2irr}) are purely geometrical
equations that, at this stage, have no relation to physics.

Studying small loops at some point of the manifold and parallelly
transporting vectors, we learn that torsion is related to a
translational misfit ($\rightarrow$ dislocations) and curvature to a
rotational (or Lorentz) misfit ($\rightarrow$ disclinations), see
\cite{Ruggiero}, for example. In this sense, torsion and curvature are
related to {\it external} groups, namely to the translation and to the
Lorentz groups. A closer discussion shows that a Riemann-Cartan
geometry can be gotten by studying the gauging of the Poincar\'e
group, the semi-direct product of the translation and the Lorentz
group. In other words, a Riemann-Cartan geometry is interrelated with
the Poincar\'e group of the tangent Minkowski space. These geometrical
facts definitely exclude the possibility to relate torsion to an {\it
  internal} group, like the U(1)-phase group of electrodynamics, for
example. A proper understanding of geometry excludes such a
possibility.

Physics is brought into the Riemann-Cartan spacetime by specifying a
gravitational Lagrange 4-form as of the Hilbert-Einstein type
according to $L_{\rm grav}=\frac{1}{2\kappa}
^{\star\!}\!\left(\vta^\a\wedge\vta^\b\right)\wedge R_{\a\b}$, where
$\kappa$ is Einstein's gravitational constant. Then we find the
following two field equations:
\begin{eqnarray}\label{field1}
\frac 12\,\eta_{\a\b\g}
\wedge R^{\b\g}&=& \kappa \Sigma_\a\,,\\
\label{field2}
\frac 12\,\eta_{\a\b\g}
\wedge T^\g&=&\kappa \tau_{\a\b}\,.
\end{eqnarray}
Here $\Sigma_\a$ is the energy-momentum 3-form of matter and
$\tau_{\a\b}$ the spin angular momentum 3-form of matter.  Upon
substitution of (\ref{field1}) and (\ref{field2}) into the irreducible
pieces of the Bianchi identities (\ref{Bianchi1irr}) and
(\ref{Bianchi2irr}), respectively, we recover the {\it
  energy-momentum} and the {\it angular momentum} laws \cite{HMcCrea}.
In this way we see again the close connection of the ECT to the
Noether theorem and the Poincar\'e group. Clearly, all this couldn't
work if torsion (or curvature) would be identified with some
electromagnetic field. Energy-momentum and angular momentum are
related to translations and Lorentz rotations and under no
circumstances to an internal group.

The ECT predicts the existence of a very weak spin-spin-contact
interaction proportional to the gravitational constant. It doesn't
show up at ordinary laboratory conditions. For vanishing material
spin, $\tau_{\a\b}=0$, one recovers general relativity.

\section{The ansatz of Evans}

Evans takes over the equations (\ref{nonmetricity}) to
(\ref{Bianchi2}) without a proper motivation. As we pointed out above,
in the framework of the ECT, the motivation lies in taking the
Poincar\'e group of Minkowski space (which yields the mass-spin
classification of elementary particles) as basis and then one gauges
this group. This was exactly what \'Elie Cartan had in mind
\cite{Cartan} when he called a Riemann-Cartan space as a space with
{\it Euclidean connection}. In the small, the Riemann-Cartan space is
just a Minkowski space. And these small Minkowski ``grains'' of a
Riemann-Cartan space are translated and Lorentz rotated with respect
to each other.

Evans instead just formally takes the equations (\ref{nonmetricity})
to (\ref{Bianchi2}) and brings in his physics by assuming that the
coframe, apart from a constant scalar factor $A^{(0)}$, is related to an
``electromagnetic potential'' by the ad hoc ansatz
\begin{equation}\label{evans1}
  A^\a=A^{(0)}\vta^\a\,.
\end{equation}
In coordinate components, we have
\begin{equation}\label{evans2}
  A_i{}^\a=A^{(0)}e_i{}^\a\,.
\end{equation}
One should compare \cite{Evans}, Eq.(12). Evans denotes the components
of the coframe by $q^a{}_\mu$. Clearly, the Evans potential $A^\a$ has
16 independent components, quite in contrast to the 4 components of
the electromagnetic potential $A$ of Maxwell's theory. Of course, the
zeroth component of $A^\a$, namely $A^0$, is {\it not} covariant under
frame transformations and cannot feature as the Maxwellian potential.
The same is true for $A^1$, $A^2$, or $A^3$ likewise. According to
Evans, we have then for the Evans field strength
\begin{equation}\label{evans3}
  F^\a=DA^\a\qquad{\rm or}\qquad F^\a
  =A^{(0)}\,D\vta^\a=A^{(0)}\,T^\a\,.
\end{equation}

(i) Apart from the geometrical arguments which I gave, namely that
torsion is related to the translation group (and not to the $U(1)$),
(ii) it is also impossible to relate the 6 components of the Maxwell
field strength 2-form $F:=dA$ (here $A$ and $F$ are the quantities in
Maxwell's theory) to the 24 components of the torsion 2-form $T^\a$:
They just don't match. And to take only one component of the torsion
$T^0=T_{ij}{}^0\,dx^i\wedge dx^j$ won't help either: It is {\it not} a
covariant relation, in spite of Evan's claim to the contrary (see
\cite{Evans}, page 10).

One could think that one decomposes the torsion $T^\a$ into its 3
irreducible pieces and attributes the Maxwellian $F$ to one of these
pieces. However, this is not possible since these pieces have 16+4+4
independent components, respectively. No irreducible piece has 6
independent components. Hence this possibility is ruled out.

``The Maxwell Heaviside theory is further restricted by the fact that
it implicitly suppresses an index $a$, meaning that only one unwritten
scalar components of the tangent bundle spacetime is considered, and
then only implicitly'' we are told (\cite{Evans}, page 10, last
paragraph). It is clear that one cannot {\it implicitly suppress} the
index $\a$ in (\ref{evans3}) (in Evan's notation the index $a$) and
then somehow get a covariant equation. This is just wishful thinking.
With the frame $e_\a$ one could build the expression $e_\a\rfloor
(DT^\a)$ and would find 6 independent components. However, we were
still left with the 24 components of $T^\a$ without being able to
reduce them in a covariant way to just 6 independent components.

The ansatz (\ref{evans1}) is not only nonsensical from a geometric
point of view, it has nothing to do with electrodynamics and with
Maxwell's theory either.

\centerline{=======}
\end{document}